\documentstyle[12pt,epsfig,epsf,feynarts,color]{article}
\def\ga{\mathrel{\raise.3ex\hbox{$>$\kern-.75em\lower1ex\hbox{$\sim$}}}}
\def\la{\mathrel{\raise.3ex\hbox{$<$\kern-.75em\lower1ex\hbox{$\sim$}}}}

\newcommand{\wt}{\widetilde}

\newcommand{\heta}{{\theta}_f}

\def\beq {\begin{equation}}
\def\eeq {\end{equation}}
\def\bec {\begin{center}}
\def\ec {\end{center}}
\def\sq {\widetilde q}
\def\a {\alpha}
\def\b {\beta}
\def\mV {m_{\!V}}

\begin{document}
\begin{center}
{\large{ {\bf Complete one--loop analysis 
         to stop and sbottom  decays 
          into $Z$ and $W^{\pm}$  bosons }}}

\vspace{.5cm}
Abdesslam Arhrib$^{1,2,3}$ and Rachid Benbrik$^2$ 
\\
\vspace{.5cm}
$1.$ National Center for Theoretical Sciences,
PO-Box 2-131 Hsinchu, Taiwan 300\\
$2.$ LPHEA, Facult\'e des Sciences-Semlalia,
B.P. 2390 Marrakech, Morocco.\\
$3.$ Facult\'e des Sciences et 
Techniques de Tanger,
B.P 416 Tanger, Morocco.
\end{center}

\begin{abstract}
We study radiative corrections to third generation scalar
 fermions into gauge bosons $Z$ and $W^\pm$. 
We include both SUSY-QCD, QED and full 
electroweak corrections. It is found that the electroweak corrections
can be of the same order as the SUSY-QCD corrections and 
interfere destructively in some region of parameter space.
The full one loop correction can reach 10\% in some SUGRA scenario,
 while in general MSSM, the one loop correction can reach 20\% for 
large 
$\tan\beta$ and large trilinear soft breaking terms $A_b$.
\end{abstract}

\noindent
{\bf 1.} Supersymmetric theories predict the  existence of scalar 
partners 
to all known quarks and
leptons. In Grand unified SUSY models, the third generation of 
scalar fermions, $\wt{t}, \wt{b}, \wt{\tau}$, gets a special status;
due to the influence of Yukawa-coupling evolution,
the light scalar fermions of the third generation are expected to be 
lighter 
than the scalar fermions of the first and second generations. 
For the same reason, the splitting between the physical masses
of the third generation may be large enough to allow the opening of the 
decay 
channels like : $\wt{f}_2 \to \wt{f}_1 V$ and/or $\wt{f}_2 \to \wt{f}_1 
\Phi $,
where $V$ is a gauge boson and $\Phi$ is a scalar boson.
\\
Until now there is no direct evidence for SUSY particles,
and under some assumptions on their 
decay rates, one can only set lower limits on their masses \cite{CDF}.
 It is expected that the next generation of $e^+e^-$ machines and/or
hadron colliders (LHC and Tevatron)
could establish the first evidence for 
the existence of SUSY particles if they are not too heavy.
\\
If SUSY particles would be detected at hadron colliders,
their properties can be studied with high accuracy at a 
high-energy linear $e^+ e^-$ collider~\cite{LC}. 
It is thus mandatory to incorporate effects beyond leading order
into the theoretical predictions, both for production and decay rate,
in order to match the experimental accuracy.\\
In this spirit, radiative corrections  to the decays of SUSY
particles have been carried out.
In particular, the QCD corrections to scalar quark decay into quarks 
plus 
charginos or neutralinos have been studied in \cite{gh1}, while
the full one loop analysis has been addressed in \cite{gh2} and found 
to 
have important impact on the partial decay widths of scalar fermions.
The QCD corrections to the decays of heavy 
scalar quarks into light scalar quarks and Higgs bosons
are found to be important \cite{DHAJ}.

Obviously, most of the studies concentrated  on the 
production and decay of light states $\wt{t}_1, \wt{b}_1$ and 
$\wt{\tau}_1$, while heavier states received less attention 
\cite{gh2,DHAJ,9701336,ah}.
These heavy states can be produced both at LHC and/or at the 
future $e^+e^-$ linear colliders.
The decay of the heavier states third generation scalar fermions
is more complicated than the light one. One can basically have four set
of two-body decays:
i) Strong decay for stop and sbottom 
$\wt{t}_2\to t \wt{g}$, $\wt{b}_2 \to b \wt{g}$ : 
if these decay are kinematically open they are the dominant one. ii) 
decay to chargino and neutralino : $\wt{f}_2\to f \wt{\chi}_i^0$, 
$\wt{f}_2\to f^\prime \wt{\chi}_i^+$.\\
If the splitting between light and heavy third generation scalar 
fermions
is large enough we may have the following decays:
iii) $\wt{f}_2\to \wt{f}_1 \Phi^0$, 
$\Phi^0=h^0,H^0,A^0$, and $\wt{f}_2\to \wt{f^\prime}_1 H^\pm$.
iv) $\wt{f}_2\to \wt{f}_1 Z^0$ and $\wt{f}_2\to \wt{f^\prime}_1 
W^\pm$.\\
\\
In the MSSM, the decay modes 
$\wt{f}_2\to \wt{f}_1 Z^0$ and  $\wt{f}_i\to \wt{f^\prime}_j W$, 
if open and under some assumptions, may be the dominant one.
Note also that in several benchmarks scenarios 
for SUSY searches, the bosonic decay of $\wt{t}_i$  and $\wt{b}_i$ 
may be the dominant \cite{SPS}. For example, in SPS5 scenario
the dominant bosonic decay have the following branching ratios 
\cite{SPS}:
$Br(\wt{b}_{1}\to W^- \wt{t}_1)=81\%$,
 $Br(\wt{b}_{2}\to W^- \wt{t}_1)=64\%$ and 
$Br(\wt{t}_{2}\to Z^0\wt{t}_1)=61\%$. While in SPS1 scenario,  
we have:  $Br(\wt{b}_{2}\to W^- \wt{t}_1)=34\%$ and 
$Br(\wt{t}_{2}\to Z^0\wt{t}_1)=23\%$.
\\
Here, we provide the complete one loop radiative 
corrections to $\wt{f}_2\to \wt{f}_1 Z^0$ and  
$\wt{f}_i\to \wt{f^\prime}_j W$
including real photon emission \cite{ab}, and discuss their effects  
in combination with the SUSY-QCD corrections \cite{qcdv}. We show that 
SUSY-QCD corrections can interfere destructively with electroweak one.\\

\noindent
{\bf 2.} The tree--level decay width for $\sq^\a_i \to \sq^\b_{j} V$ 
can thus be written as:
\beq
\Gamma^{0} (\sq^\a_i \to \sq^\b_{j} V) =
\frac{  ( g_{V \wt{f}_i \wt{f}_j} )^2\, \kappa^3 (m_i^2, m_j^2, \mV^2)
}{ 16\pi\, \mV^2\, m_i^3 } ,
\label{tree}
\eeq
with $\kappa (x,y,z) = (x^{2}+y^{2}+z^{2}-2xy-2xz-2yz)^{1/2}$,
$g_{Z\wt{f}_1 \wt{f}_2}= \frac{e}{s_Wc_W}  \{ 
(I_3^f -Q_f s_W^2)\sin\theta_f\cos\theta_f +   
Q_f s_W^2 \cos\theta_f \sin\theta_f$,\,\,$g_{Z\wt{f}_1 \wt{f}_2}= g_{Z\wt{f}_2 \wt{f}_1}$ and $g_{W{\widetilde{f}_1}\widetilde{f^\prime}_{2}} =
\frac{e}{\sqrt{2}s_{W}}\cos\theta_f\sin\theta_{f'}$, $I_3^f$ is the isospin.\\
\\
At one loop level,  the decay  $\sq^\a_i \to \sq^\b_{j} V$
receive contributions from vertex diagrams, 
gauge boson $V$ and scalar fermions $\sq^\a_i$ self energies
as well as theire mixings (see \cite{ab} for details).
Note that the transitions between gauge bosons and scalar bosons
like $W^\pm$-$H^\pm$, $W^\pm$-$G^\pm$, $Z^0$-$A^0$, $Z^0$-$G^0$ are 
present.
Owing to Lorentz invariance, those mixing are proportional to 
$p_V^\mu$ momentum; then since the vector gauge bosons $W$ and $Z$ are 
on-shell transverse, those transitions vanish. In what follows we will
ignore vector-scalar boson mixing.
\\
We have evaluated the one-loop amplitudes in
the 't Hooft--Feynman gauge using FeynArts and FormCalc~\cite{FA,FF}. 
The one-loop amplitudes are ultraviolet (UV) 
and infrared (IR) divergent. 
The UV singularities are treated by  dimensional
reduction and are compensated
in the on-shell renormalization scheme. 
We have checked explicitly   
that the results are identical in using dimensional
reduction and dimensional regularization. 
The IR singularities are regularized with a 
small fictitious photon mass $\delta$ and are 
eliminated by adding to the one loop contribution both
real-photon and real-gluon emission \cite{ab},
$\wt{f}_i\to  \wt{f}_j^* V \gamma$ and $\wt{f}_i\to  \wt{f}_j^* V g$.

Recently, there have been several developments
in the renormalization of MSSM. Several schemes 
are available \cite{ren}.
Here, we follow the strategy of~\cite{gh2,ah} by introducing 
counter-terms for the physical parameters, i.e. for masses and
mixing angles, and perform  field renormalization
in a way that residues of renormalized propagators 
can be kept at unity.\\
For SM parameters and fields, we will adopt throughout, 
the on--shell renormalization scheme of Refs. \cite{Denner}. 
In the on--shell scheme we use the mixing angle $s_W$ (resp $c_W$) 
is defined by  $s_W^2=1-M_W^2/M_Z^2$ (resp $c_W^2=M_W^2/M_Z^2$). 
Its counter--term  is completely fixed by the mass 
counter-terms of W and Z gauge bosons.

The extra parameters and fields we still have to renormalize in our 
case
are the scalar fermion wave functions $\wt{f}_i$ and the mixing angle
${\theta}_{f}$ which enter in the tree level amplitude 
eq.~(\ref{tree}).
\\
In the general case, where sfermions mixing is allowed, 
the wave functions of the two sfermions mass eigenstates are not 
decoupled. 
Taking into account the mixing, the renormalization of the
sfermions wave functions and the mixing angle 
$\heta$ can be performed by making the
following substitutions in the Lagrangian 
\begin{eqnarray}
\widetilde{f}_1 \rightarrow Z_{11}^{1/2} \wt{f}_1+
Z_{12}^{1/2} \wt{f}_2 \ \ , \ 
\wt{f}_2 \rightarrow Z_{22}^{1/2} \wt{f}_2 + Z_{21}^{1/2} \wt{f}_1 \ \
, \ \ 
\heta \rightarrow \heta+\delta \heta  \label{cd3}
\end{eqnarray}

To fix all the renormalization constants,
we use the following renormalization conditions:\\
i) The on-shell conditions for $m_W$, $m_Z$, $m_e$ 
and the electric charge $e$ are defined as in the Standard 
Model \cite{Denner}.\\
ii) On-shell condition for the scalar fermion $\widetilde{f}_i$ :
we choose to identify
the physical scalar fermion mass with the corresponding parameter in
the renormalized Lagrangian,
and require the residue of the propagators to have its tree-level 
value, i.e., 
\begin{eqnarray}
\delta Z_{ii} & = & -\Re\{\frac{\partial}{\partial p^2}(
{\Sigma}_{\wt{f}_i \wt{f}_i} (p^2))\} |_{p^2=m^2_{\wt{f}_i} }   
 \  , \   
\delta Z_{ij}=
\frac{\Re\{{{\Sigma}}_{ \wt{f}_i \wt{f}_j }(m_{\wt{f}_j}^2)\}}
{m_{\wt{f}_j}^2-m_{\wt{f}_i}^2} \ , \
\delta
m^2_{\wt{f}_i}   =   \Re ({\Sigma}_{\wt{f}_i \wt{f}_j}
(m^2_{\widetilde{f}_i}))
\end{eqnarray}
where $\sum_{\wt{f}_i \wt{f}_j } (p^2)$, $i,j=1,2$ is the scalar 
fermion bare self-energy.\\
iii) For the renormalization condition which defines the mixing
angle $\heta$, we select this condition in such a way to kill the 
transitions $\wt{f}_i \leftrightarrow \wt{f}_j$ at the one--loop level.
The renormalization of the scalar fermion mixing 
angle is then given by \cite{gh2}:
\begin{eqnarray}
\delta{\theta}_{{f}} = \frac{1}{2} \frac{\Sigma_{\wt{f}_i \wt{f}_j}(
m^2_{ \wt{f}_j} )+ \Sigma_{\wt{f}_i \wt{f}_j}(m^2_{ \wt{f}_i} ) }
{ m^2_{ \wt{f}_j} - m^2_{ \wt{f}_i} } \label{angle}
\end{eqnarray}

\noindent
{\bf 3.} Now we are ready to 
present our numerical results both for the tree-level and one-loop 
decay widths and branching ratios for $\tilde{f}_i \to \tilde{f}_j Z$ 
and
$\tilde{f}_i \to \tilde{f'}_j W^{\pm}$. 
Let us first fix our inputs and SUSY parameters choice.\\
As experimental data points, the following input quantities enter:
$\alpha^{-1}=137.03598$, $m_Z=91.1875$ GeV, 
$m_W=80.45$ GeV. For quarks masses, we use
effective quark masses that reproduce the hadronic vacuum polarization
contribution $\Delta \alpha(m_Z^2)$ with a sufficiently high accuracy
have been chosen  \cite{Eidelman}.
\\
For the SUSY parameters, we will use MSSM inputs which look like
some of  the Snow-mass Points and Slopes (SPS) and 
benchmarks scenarios for SUSY searches \cite{SPS}.
For our study we will use SPS1 and SPS5 scenario. As we explained
in the introduction, for those 2 scenarios the bosonic decays 
of scalar fermions $\wt{f}_i \to  \wt{f}_j V$, when open, are dominant.
\begin{figure}[t!]
\smallskip\smallskip 
\vskip-1.3cm
\centerline{
{\epsfxsize3. in\epsffile{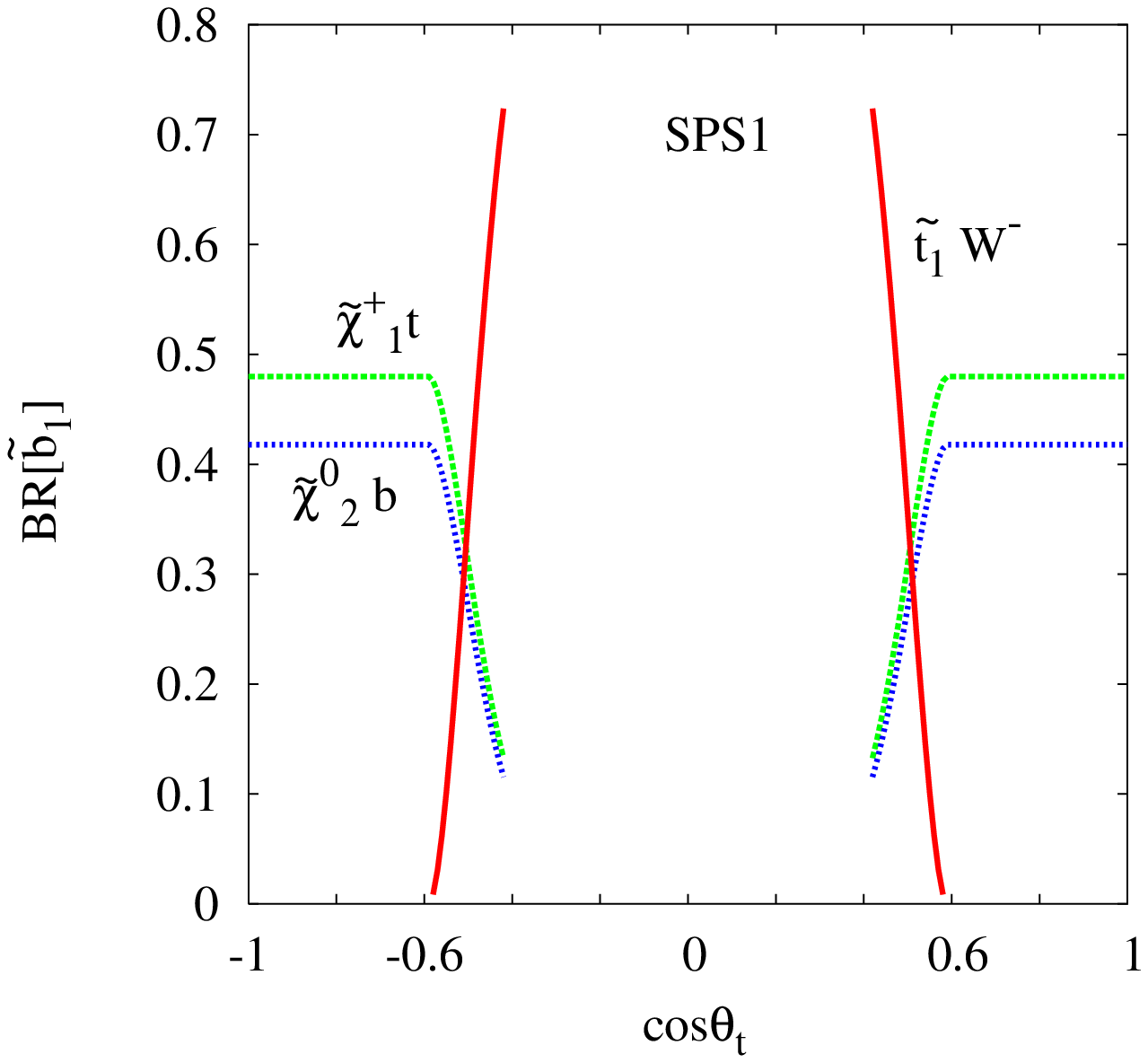}}  
\hskip-1.99cm
{\epsfxsize3.01 in\epsffile{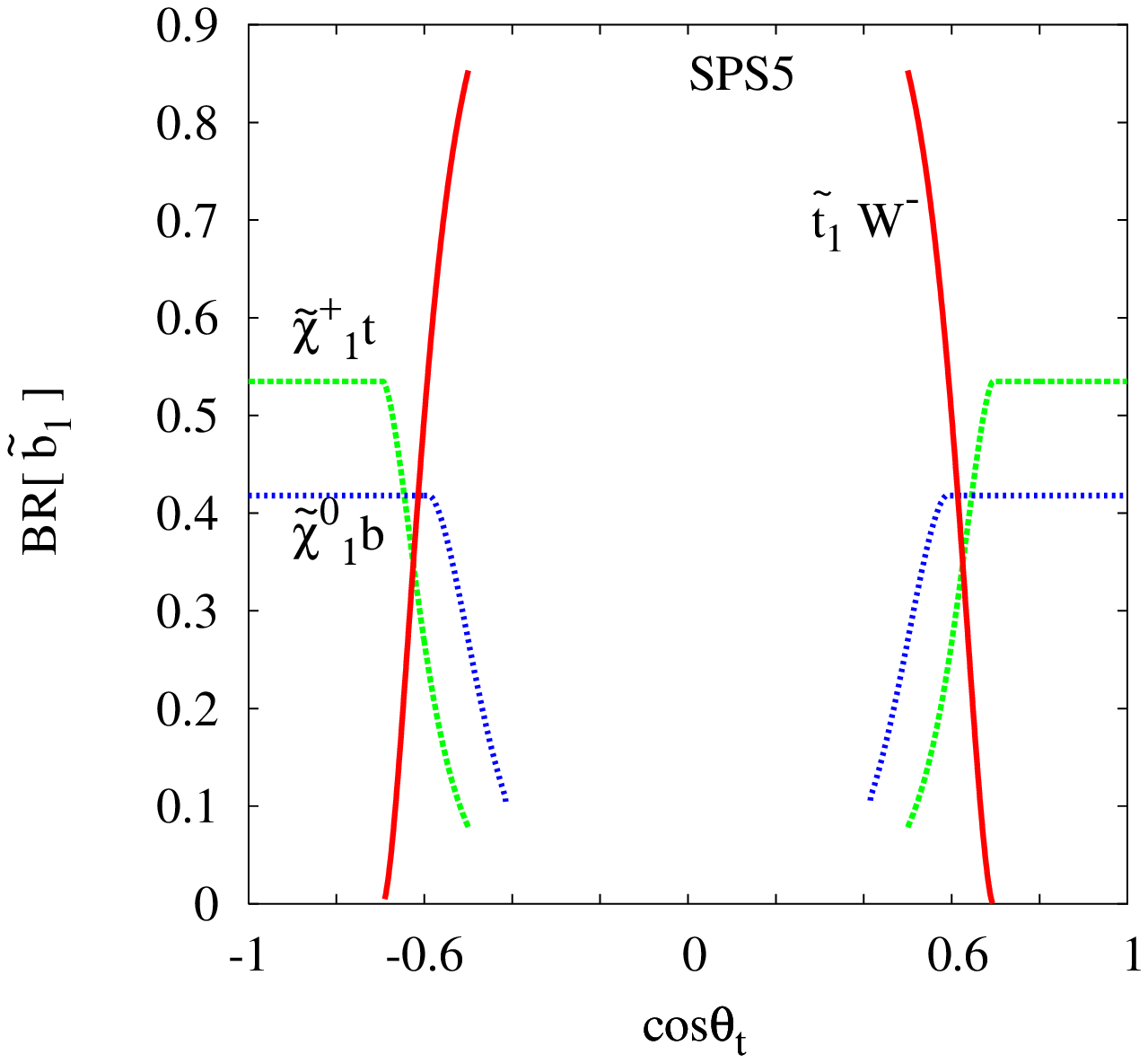}}}
\hskip-1.4cm
\smallskip\smallskip
\vskip-1.5cm
\centerline{
{\epsfxsize3. in\epsffile{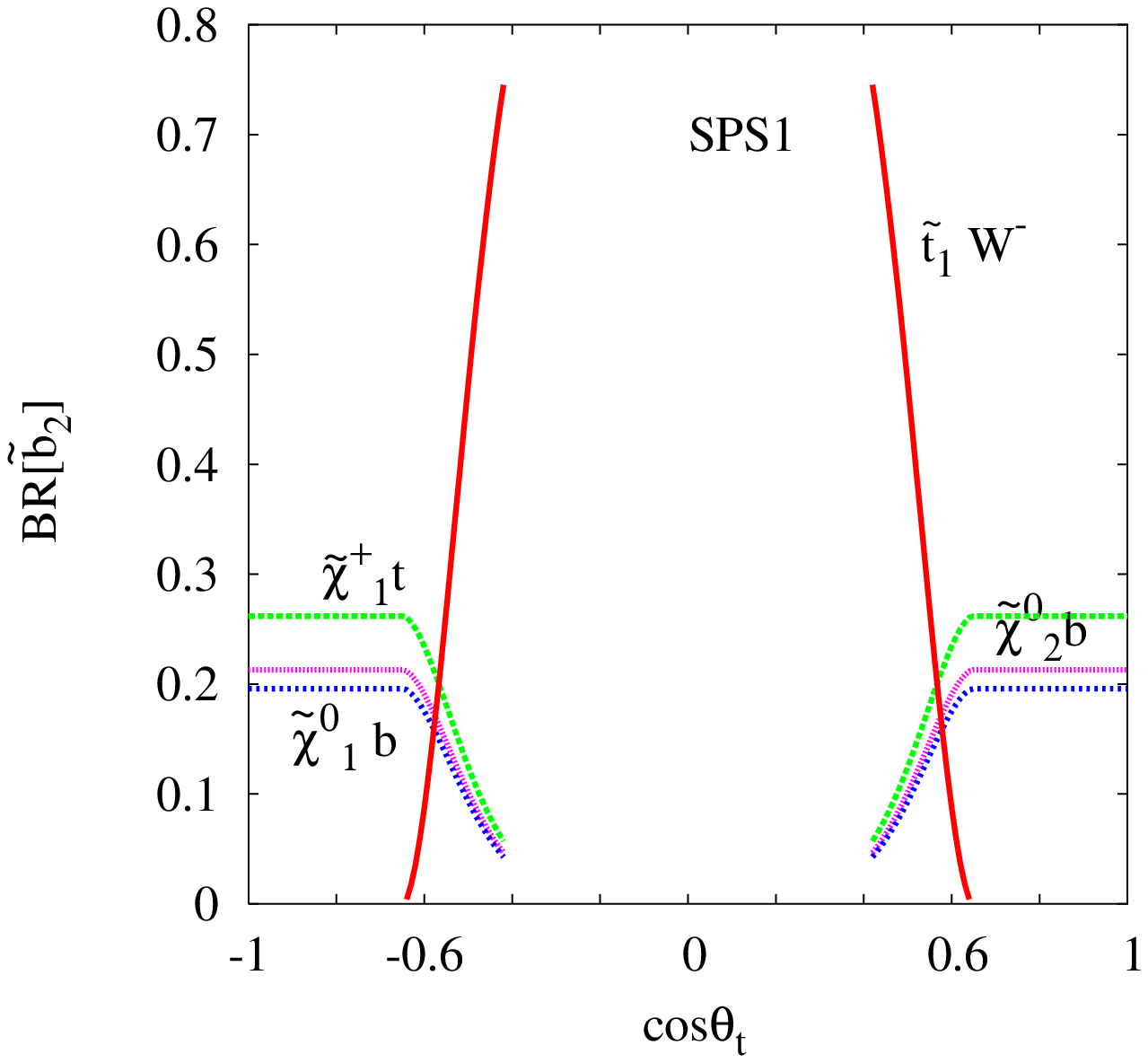}}  
\hskip-1.99cm
{\epsfxsize3.01 in\epsffile{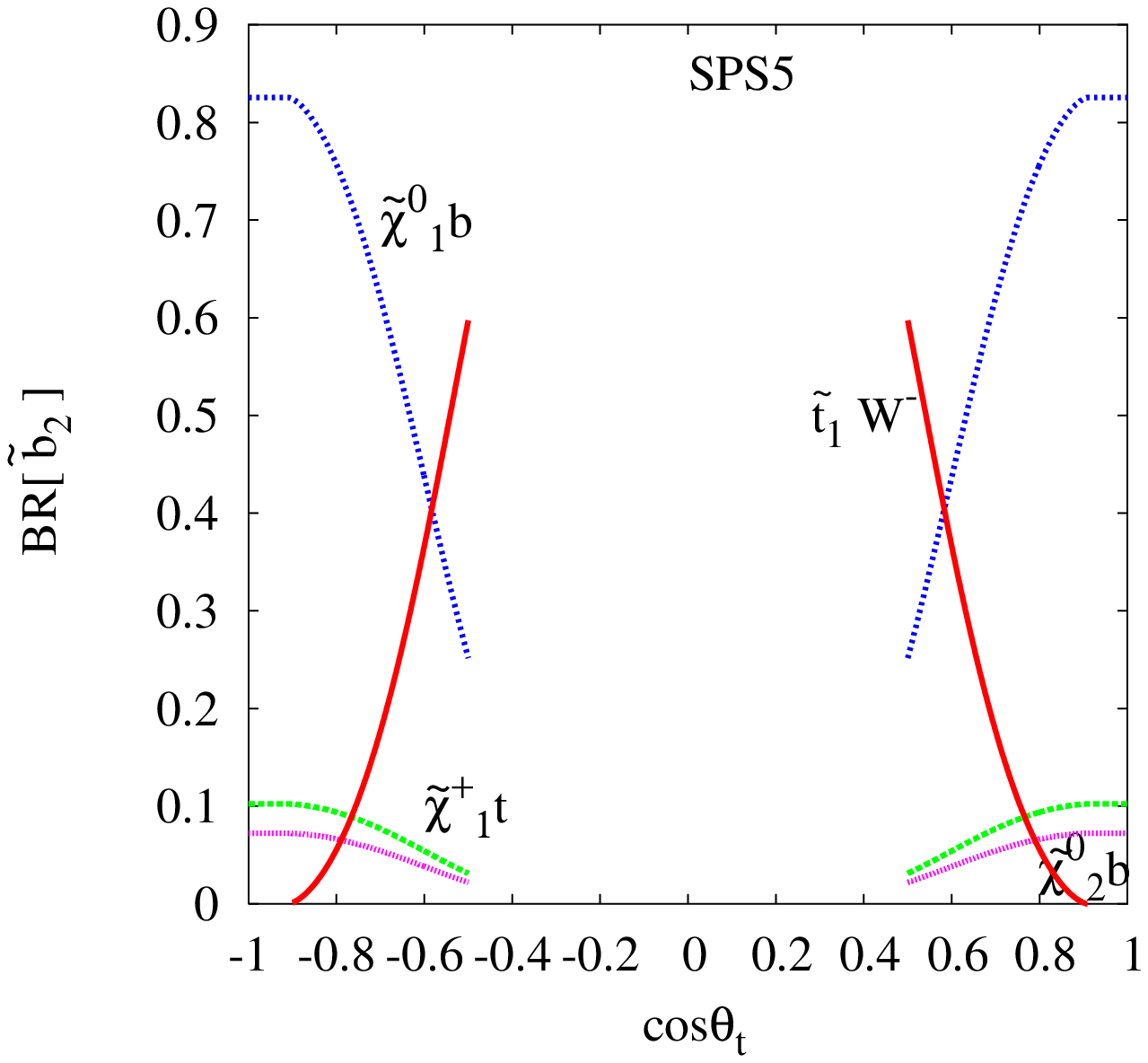}}}
\hskip-1.4cm
\smallskip\smallskip
\vskip-1.5cm
\centerline{
{\epsfxsize3. in\epsffile{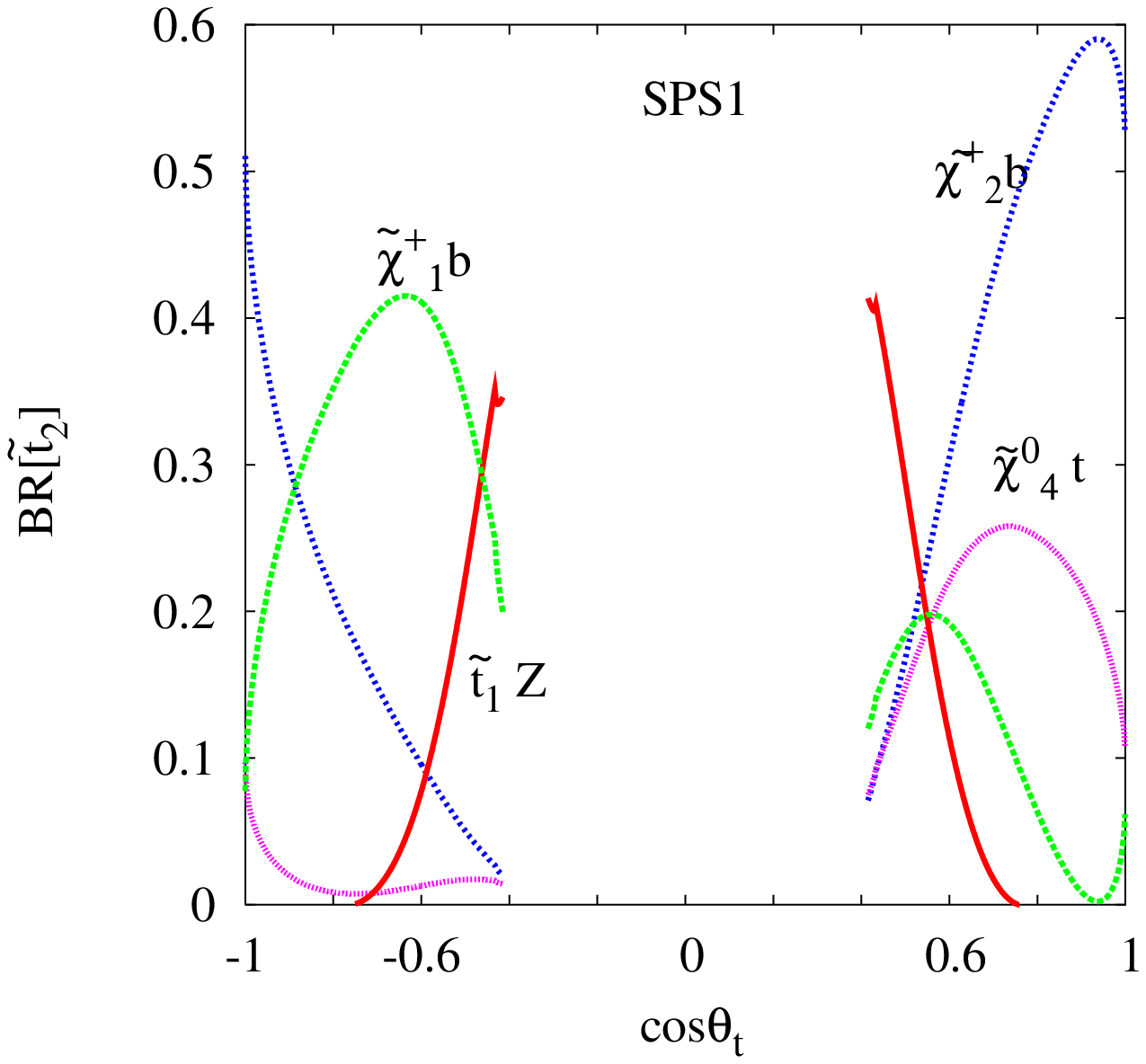}}  
\hskip-1.99cm
{\epsfxsize3.01 in\epsffile{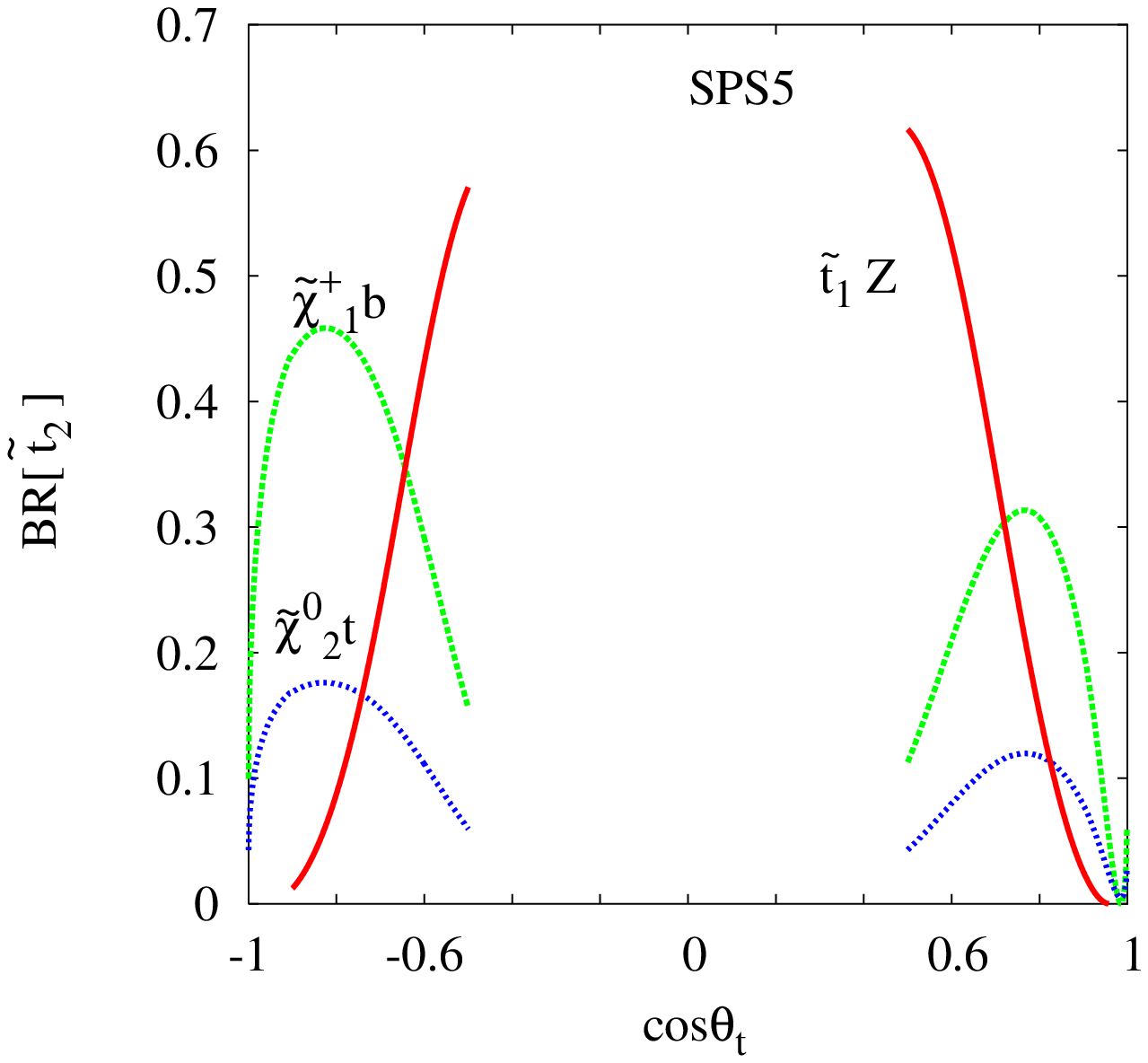}}}
\smallskip\smallskip
\vskip-.6cm
\caption{Branching ratios of bosonic decays of $\wt{b}_1$ (upper
  plots), $\wt{b}_2$ (middle plots)
and $\wt{t}_2$ (lower plots) in SPS1 (left) and SPS5 (right) as 
function of $\cos\theta_t$.  }
\label{fig4}
\end{figure}

In SPS1, we have the following spectrum (are listed only the
parameters needed):
 $\tan\beta=10$, $m_{A^0} = 394$ GeV, 
$A_t = -431.34$ GeV, $A_b =582.67 $ GeV, 
$M=193$ GeV, $M'=99$ GeV, $\mu=352$ GeV. The mass of the first and 
second generation scalar fermion is of the order 177 GeV (average).
While the masses of the third generation scalar fermions are :
$m_{\wt{t}_{1}} = 396.43$ GeV, $m_{\wt{t}_{2}} = 574.71$ GeV ,
$m_{\wt{b}_{1}} = 491.91$ GeV ,  $m_{\wt{b}_{2}} = 524.59$ GeV.
The mixing angle are given by 
$\cos\theta_t = 0.57$ , $\cos\theta_b = 0.88$.

In SPS5 (light stop scenario), we have the following spectrum :
$m_{A^0} =  694$ GeV, $\tan\beta=5$
$A_t = -785.57$ GeV, $A_b = -139.11$ GeV, 
$M=235$ GeV, $M'=121$ GeV, $\mu= 640 $ GeV
The mass of the first and second generation scalar 
fermion is of the order 231 GeV (average).
The masses of the  third generation are
$m_{\tilde t_{1}} = 253.66$ GeV , $m_{\tilde t_{2}} = 644.65$ GeV ,
$m_{\tilde b_{1}} = 535.86$ GeV ,
$m_{\tilde b_{2}} = 622.99$ GeV.
The mixing angle are given by
$\cos\theta_t = 0.54$ , $\cos\theta_b = 0.98$.

In fact, our strategy is the following :
the SPS1 and SPS5 outputs are fixed as above,
but we will allow a variation of the mixing angles $\cos\theta_t$, 
$\cos\theta_b$ from their SPS values.
According to our parametrization defined in section 2,
we choose as independent parameters
$m_{\wt{t}_2},  m_{\wt{b}_1} m_{\wt{b}_2}, {\theta}_t, {\theta}_b$ 
together with $\mu$ and $\tan\beta$. $m_{\wt{t}_1}$ and $A_f$ are fixed 
by 
\begin{eqnarray}
&&m_{\wt{t}_1}^2=\frac{1}{\cos^2 {\theta}_t}
(  m_W^2\cos 2\beta  -m_{\wt{t}_2}^2 \sin^2 {\theta}_t + 
m_{\wt{b}_2}^2 \sin^2 {\theta}_b + m_{\wt{b}_1}^2 
\cos^2 {\theta}_b +m_t^2 - m_b^2)\label{mass1}\\
&&A_f=\mu (\tan\beta)^{-2I_f} + \frac{m_{\wt{f}_1}^2 - 
m_{\wt{f}_2}^2}{m_f} 
\sin {\theta}_f\cos{\theta}_f \label{af}
\end{eqnarray}
The variation of $\cos\theta_t$
and $\cos\theta_b$ imply the variation of $m_{\wt{t}_2}$ 
as well as $A_t$ and  $A_b$.
Since we allow  variation of the
$\cos\theta_t$ and $m_{\wt{t}_2}$ mass, our inputs 
can be viewed as a general MSSM inputs and not as SPS one.\\ 
As outlined in section~2, $A_{t,b}$ are fixed by tree level relation  
eq.~(\ref{af}). Of course, $A_{t,b}$ receive radiative corrections 
at high order. However, $A_{t}$ and $A_b$ enter game only at one-loop 
level
in our processes, radiative corrections to $A_{t}$ and $A_b$ is 
considered
as two-loop effects. 
\\
In Fig.~(\ref{fig4}) we show branching ratios of 
$\wt{b}_{1}$, $\wt{b}_{2}$ and $\wt{t}_{2}$.
We evaluate the bosonic decays : 
$\wt{b}_{1}\to W^- \wt{t}_1$,
$\wt{b}_{2}\to W^- \wt{t}_1$ and 
$\wt{t}_{2}\to Z^0\wt{t}_1$ 
as well as the fermionic decays  
$\wt{f}_{i}\to \chi_i^{0} f (\chi_i^{+} {f^\prime})$
as function of $\cos\theta_t$ for SPS1 (left) and SPS5 (right) 
scenario.
From those plots, it is clear that the bosonic decay,
once open, are the dominant
one for $|\cos\theta_t|\approx 0.4\to 0.45$. For $|\cos\theta_t|\approx 
0.4$ 
the light stop $m_{\wt{t}_1}$ is about $100$ GeV, when
$|\cos\theta_t|$ 
increases, the  $m_{\wt{t}_1}$ increases also and for large 
$|\cos\theta_t|$ 
the bosonic decays are already close and the branching ratio vanishes.

We note that in the case of SPS1 
the bosonic decays are open only for $0.4\la |\cos\theta_t| \la 0.6$ 
Fig~.(\ref{fig4}) (left). 
In the region $|\cos\theta_t| \la 0.4$, the light stop is below
the experimental upper limit $m_{\wt{t}_1}\approx 90 $ GeV, and no data
are shown. While in the case of SPS5 Fig~.(\ref{fig4}) (right), for 
$|\cos\theta_t| \la 0.5$,
we find that $m_{\wt{t}_1}$ is below the experimental upper limit
and also $\delta\rho \ga 0.001$ due to large splitting
between stops and sbottoms.

The magnitude of SUSY radiative corrections can be described by
the relative correction which we define as:
\begin{eqnarray}
\Delta = \frac{\Gamma^{\rm{1-loop}}(\wt{f}_i \to \wt{f}_j V)-
\Gamma^{\rm{tree}}(\wt{f}_i \to \wt{f}_j V)}
{\Gamma^{\rm{tree}}(\wt{f}_i \to \wt{f}_j V)}\label{del}
\end{eqnarray}
\begin{figure}[t!]
\smallskip\smallskip 
\vskip-1.cm
\centerline{{
\epsfxsize3.2 in 
\epsffile{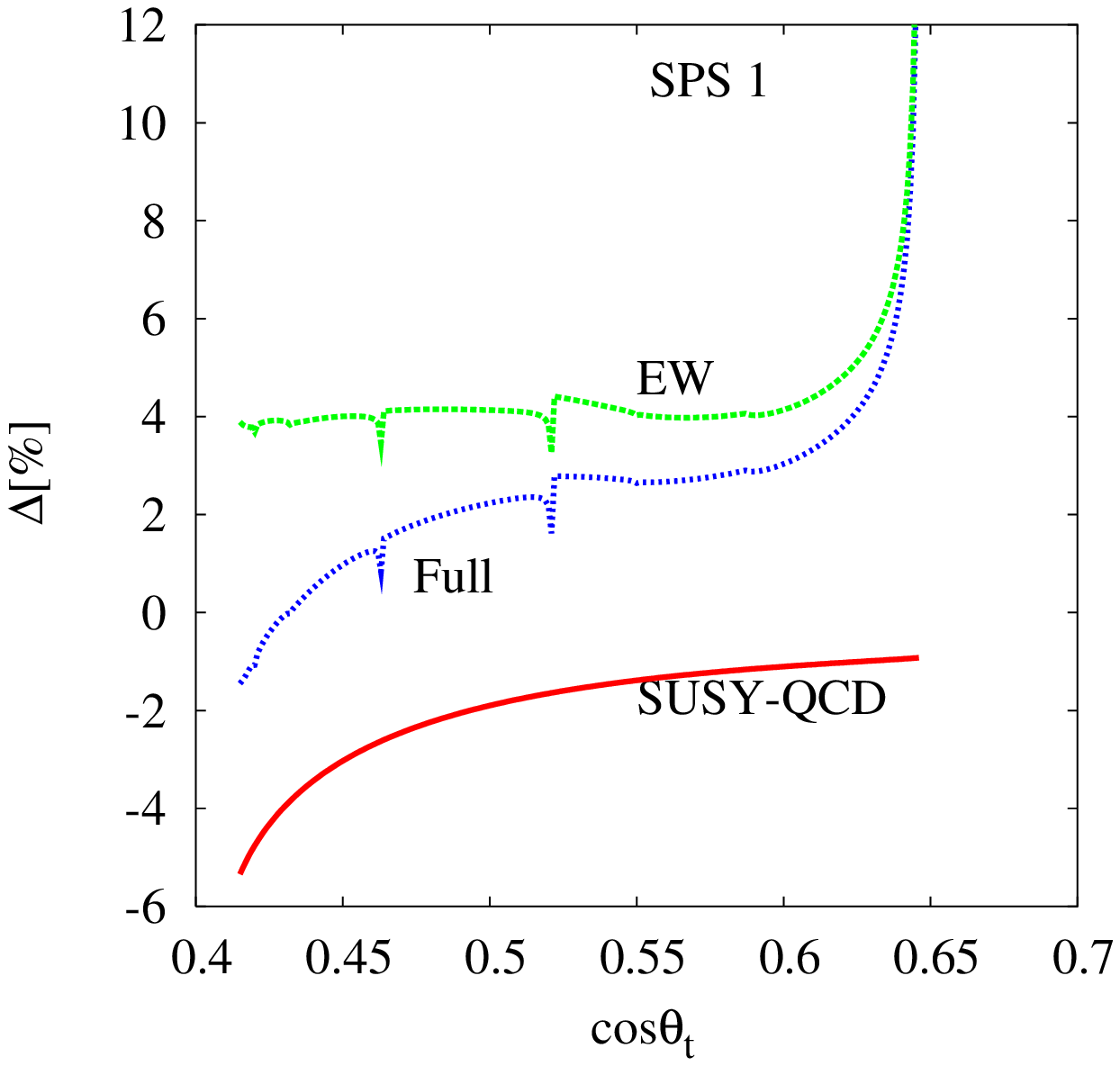}}  
\hskip-1.99cm
\epsfxsize3.2 in 
\epsffile{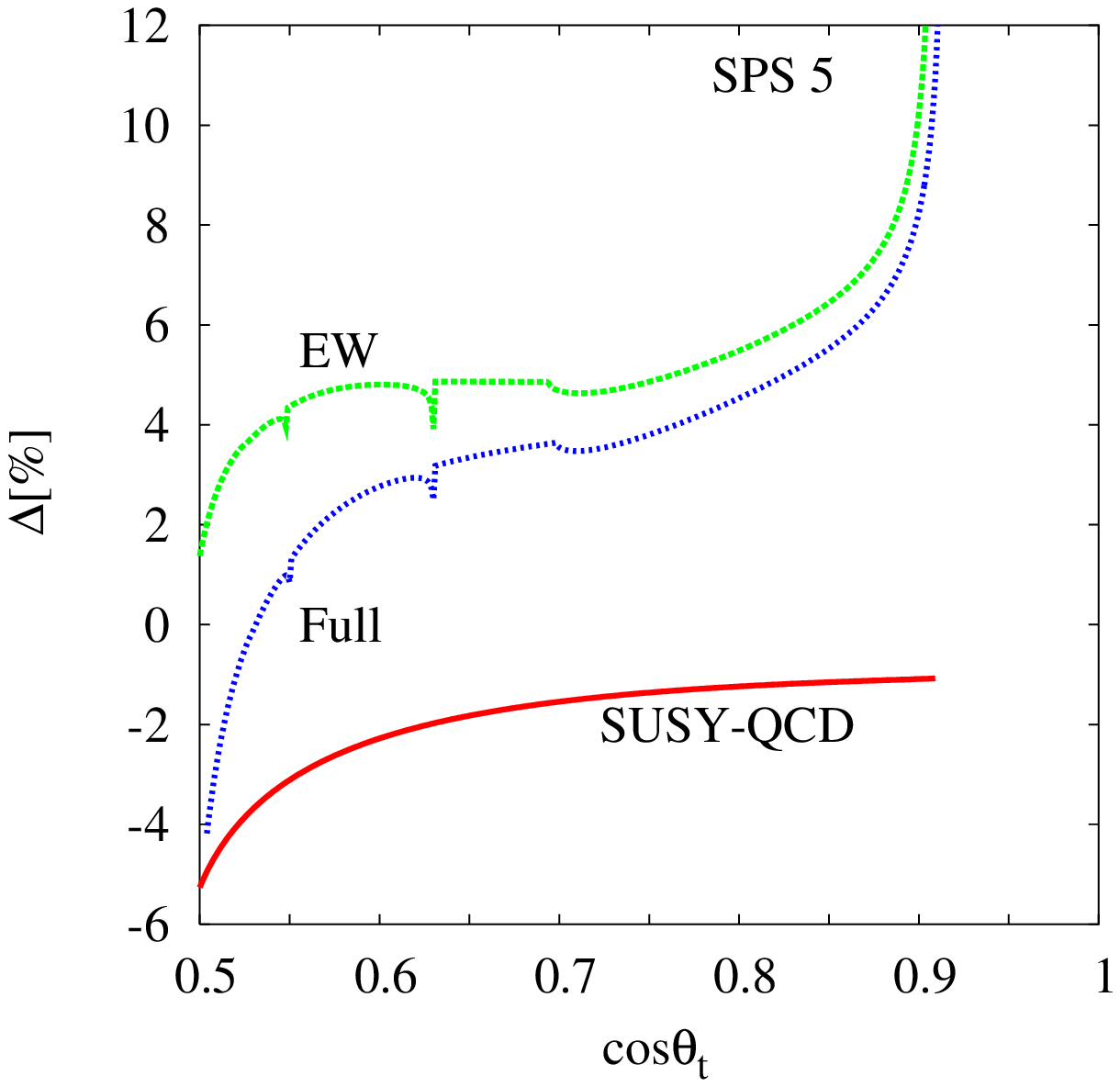} }
\smallskip\smallskip
\vskip-.40cm
\caption{Relative correction (electroweak EW, SUSY-QCD and full) 
to $\wt{b}_2 \to \wt{t}_1 W$ 
as function of $\cos\theta_t$ in SPS1 (left) and SPS5 (right)}
\label{fig5}
\end{figure}
\begin{figure}[t!]
\smallskip\smallskip 
\vskip-1.cm
\centerline{{
\epsfxsize3.2 in\epsffile{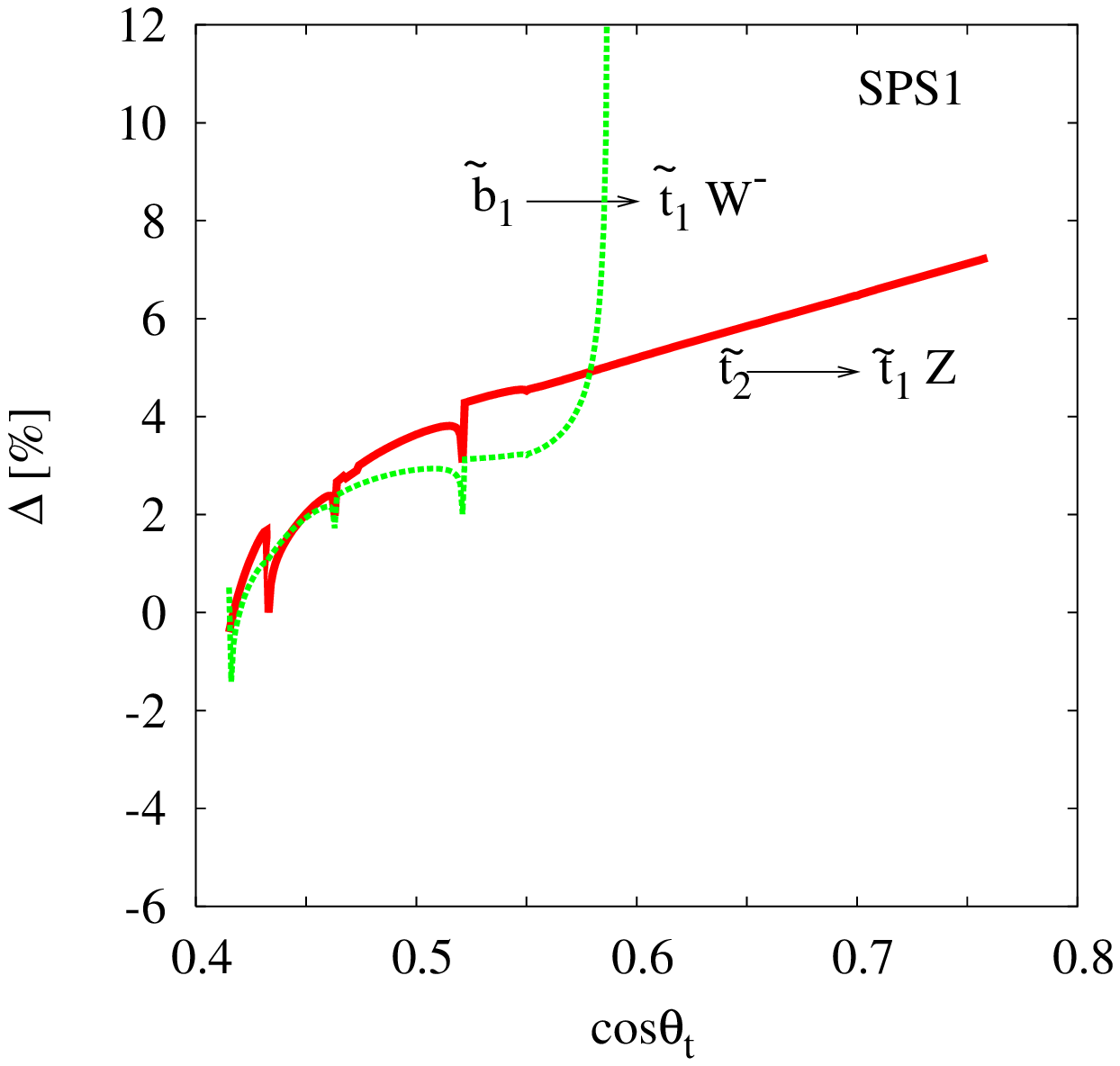}}  
\hskip-1.99cm
\epsfxsize3.2 in 
\epsffile{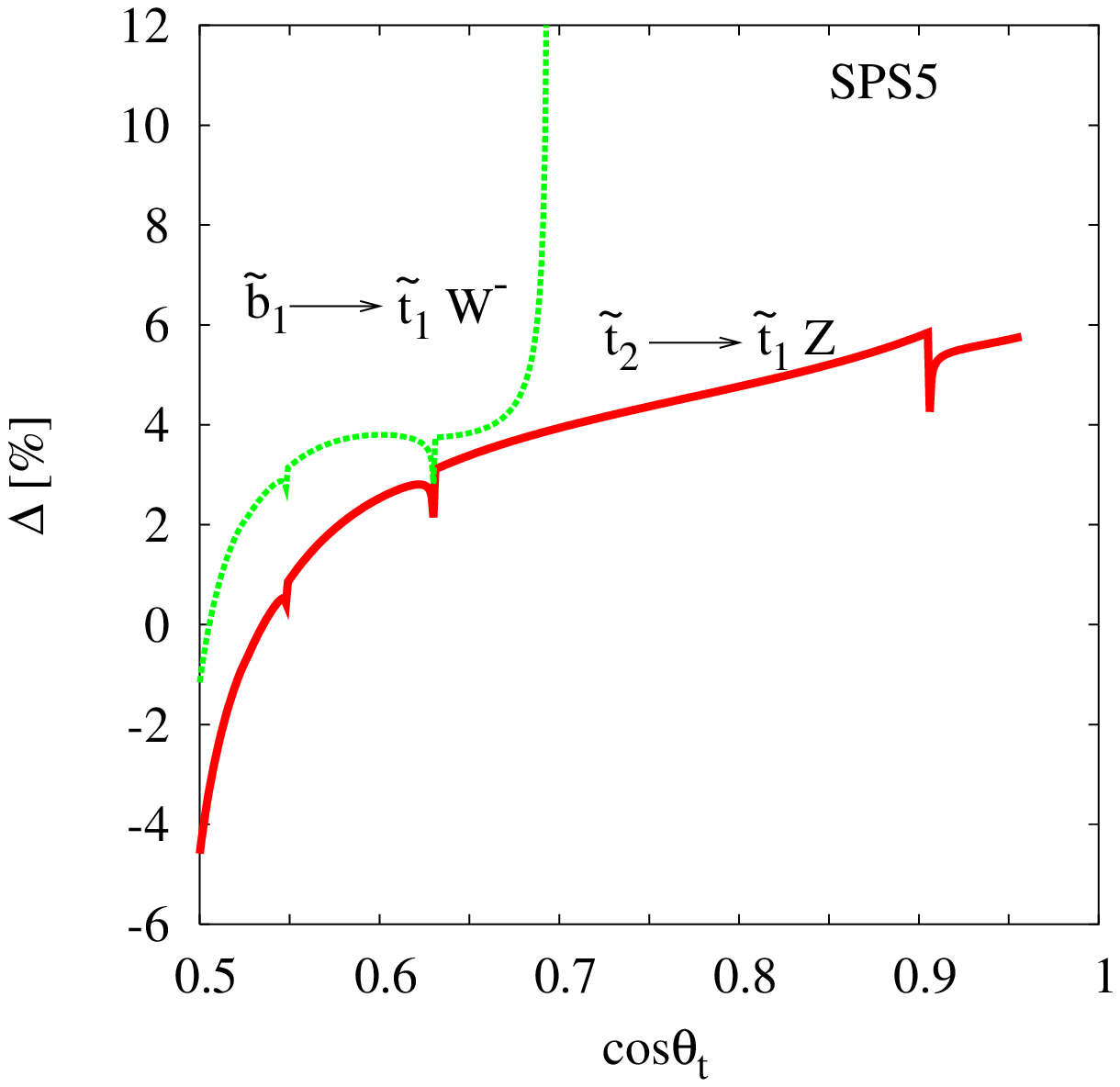}}
\smallskip\smallskip
\vskip-.40cm
\caption{Relative correction to $\wt{b}_1 \to \wt{t}_1 W$ and 
 $\wt{t}_2 \to \wt{t}_1 Z$ 
as function of $\cos\theta_t$ in SPS1 (left) and SPS5 (right)}
\label{fig55}
\end{figure}

In Fig.~(\ref{fig5}) we illustrate the relative correction
$\Delta$ as function of $\cos\theta_t$ for the decay
$\wt{b}_2 \to W \wt{t}_1$ in SPS1 (left) and SPS5 (right).
As it can be seen from the left plot, the SUSY-QCD corrections \cite{susyqcd} lies in
the range $-1\% \to -6\%$ while the EW corrections lie in the range
$4\% \to 10\%$ for $\cos\theta_t \approx 0.4 \to 0.65$.
The SUSY-QCD and EW corrections are of opposite sign,
there is a destructive interference and so the full one-loop 
corrections lie between them.
For $\cos\theta_t\to 0.65$,
the total correction increases to about 10\%. 
This is due to the fact that for $\cos\theta_t\to 0.65$
the mass of light stop is $m_{\wt{t}_1}\approx 444$ GeV, the decay 
$\wt{b}_2 \to W \wt{t}_1$ is closed and so the tree level 
width decreases to zero. 
The observed peaks around $\cos\theta_t\approx 0.46$ 
(resp $\cos\theta_t\approx 0.53$)
correspond to the opening of the transition $\wt{t}_1\to \chi_1^0 t$
(resp $\wt{t}_1\to \chi_2^0 t$).
The right plot of Fig.~(\ref{fig5}) in SPS5 scenario, 
exhibits almost the same behavior as the left plot. 
The electroweak corrections interfere destructively with the SUSY-QCD
ones, the full corrections
are between $-4\% \to 10\%$ for $\cos \theta_t\in[0.5,0.9]$.

In Fig.~(\ref{fig55}) we show the relative correction
$\Delta$ as function of $\cos\theta_t$ for the decay
$\wt{b}_1 \to W \wt{t}_1$ and  
$\wt{t}_2 \to Z \wt{t}_1$  in SPS1 (left) 
 and SPS5 (right) scenario.\\
In the case of $\wt{t}_2 \to Z \wt{t}_1$,
the total correction lies in $-1\to  7\%$ (resp $-5 \to 6\%$) 
in SPS1 (resp SPS5) scenario. 
From Fig.~(\ref{fig55}), one can see that 
the relative corrections for $\wt{b}_1 \to W \wt{t}_1$
are enhanced for $\cos\theta_t\approx 0.6$ (resp $\cos\theta_t\approx
0.75$) in SPS1 (resp SPS5). This behavior has the same explanation as 
for $\wt{b}_2 \to W \wt{t}_1$ in figure.~(\ref{fig5}).
At $\cos\theta_t\approx 0.6$ (resp $\cos\theta_t\approx
0.75$) in SPS1 (resp SPS5), the decay channel 
$\wt{b}_1 \to W \wt{t}_1$ (resp 
$\wt{t}_2 \to Z \wt{t}_1$) is closed and so the tree level 
width decreases to zero.
The observed peaks around $\cos\theta_t\approx 0.46$ (resp 
$\cos\theta_t\approx 0.53$)
correspond to the opening of the transition 
$\wt{t}_1\to \chi_1^0 t$ (resp $\wt{t}_1\to \chi_2^0 t$).\\
In all cases, we have isolated the 
QED corrections (virtual photons and real photons), we have checked 
that this contribution is very small, less than about 1\%.\\

\begin{figure}[t!]
\smallskip\smallskip 
\vskip-.5cm
\centerline{
{\epsfxsize3.2 in\epsffile{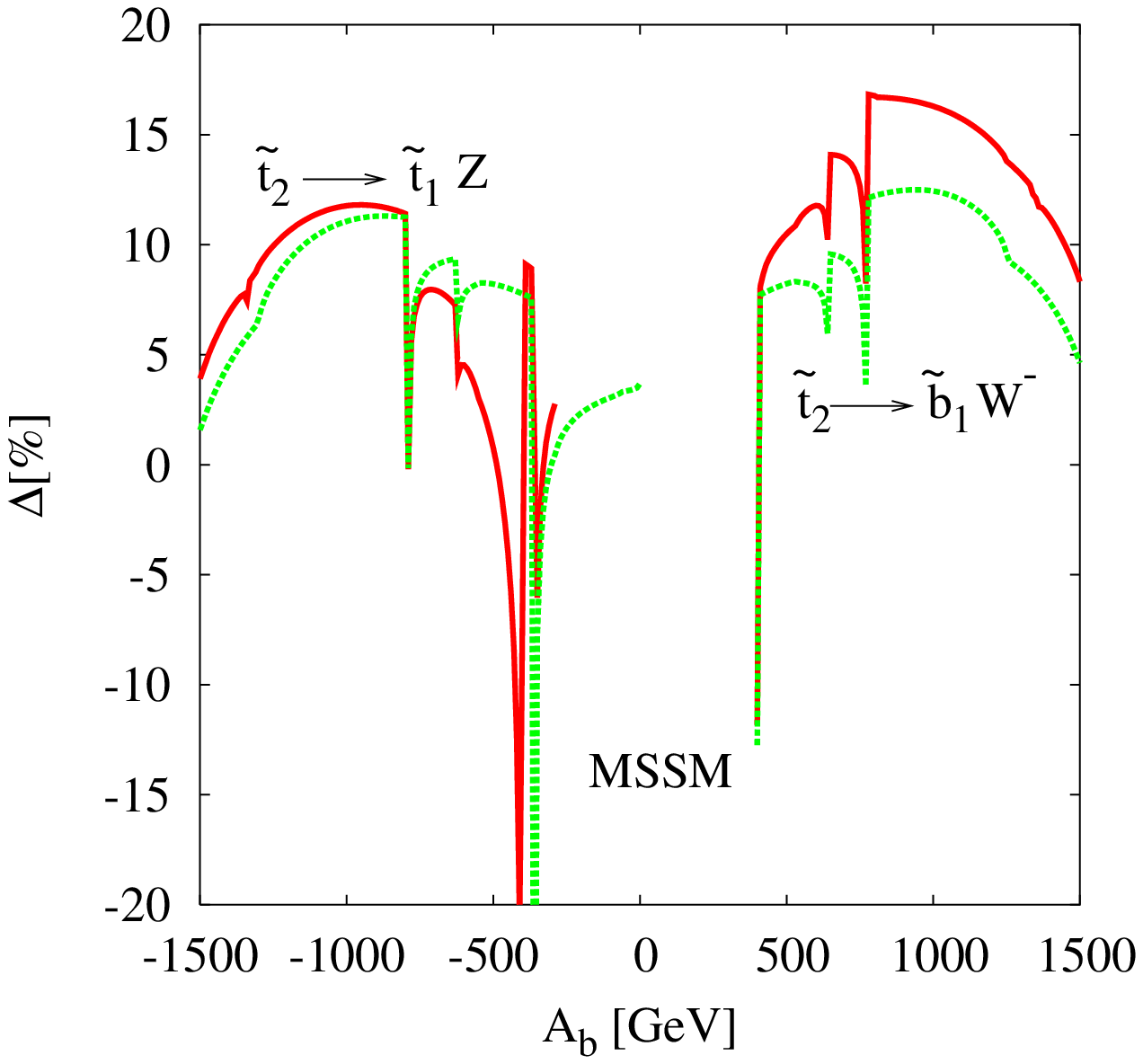}}  
\hskip-1.99cm
{\epsfxsize3.2 in\epsffile{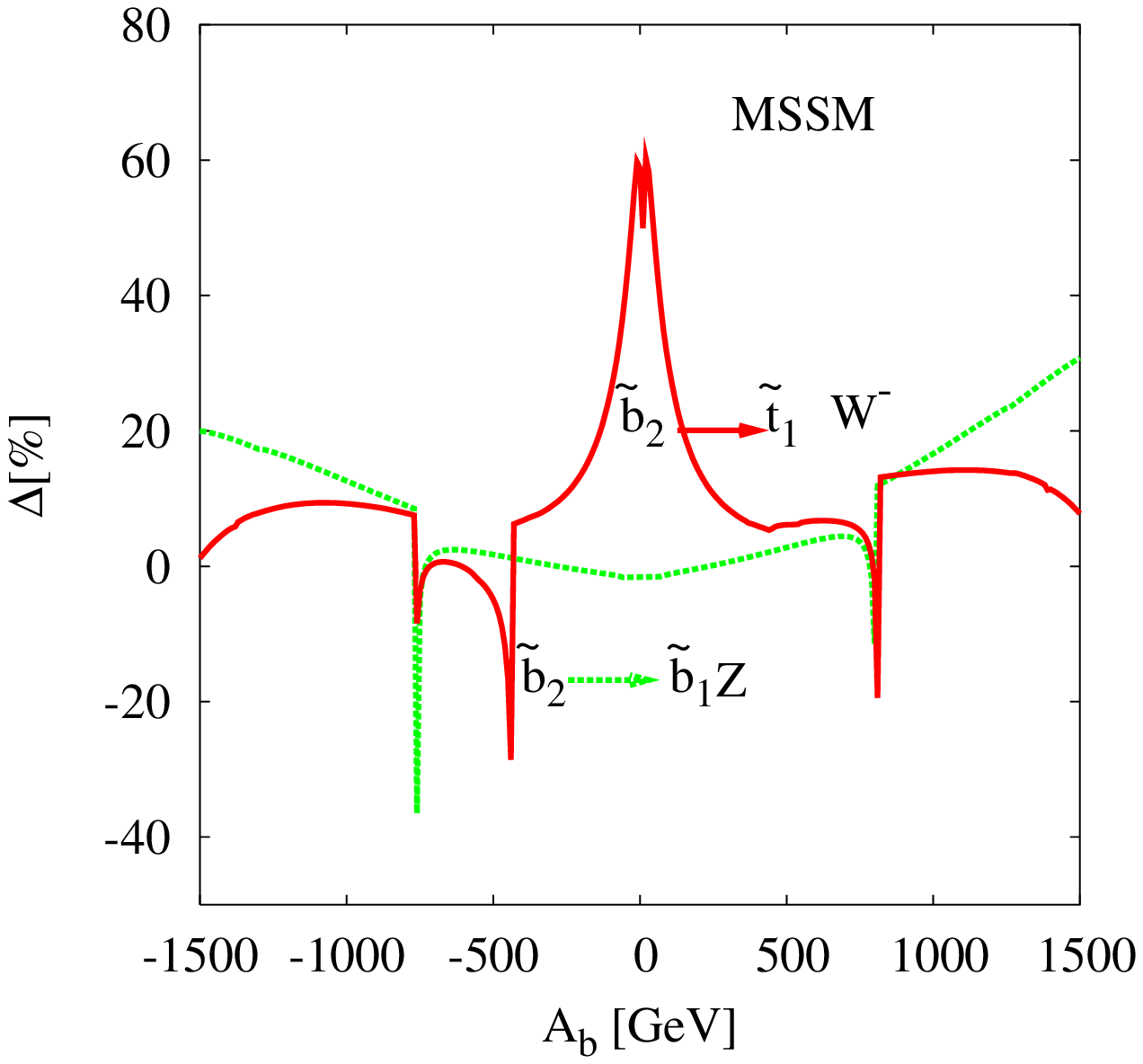}}}
\smallskip\smallskip
\caption{Relative correction to 
$\wt{t}_2 \to \wt{b}_1 W$,$\wt{t}_2 \to \wt{t}_1 Z$ (left) and $\wt{b}_2 \to \wt{b}_1 Z$,$\wt{b}_2 \to \wt{t}_1 W$ (right) as function of $A_t=A_b$ in  MSSM for $\mu,M_2,M_A=500,130, 200$ GeV and $\tan\beta=60$}
\label{mssm}
\end{figure}

Fig.(\ref{mssm}) illustrates the relative corrections 
to $\wt{t}_2 \to \wt{b}_1 W$, $\wt{t}_2 \to \wt{t}_1 Z$ (left)
and  $\wt{b}_2 \to \wt{b}_1 Z$, $\wt{b}_2 \to \wt{t}_1 W$ (right)
 as function of $A_b=A_t$ in general MSSM for 
large $\tan\beta=60$, $\mu=500$ GeV, $M_2=130$
 GeV and $M_A=200$ GeV.
It is clear from this plot that the relative corrections are bigger 
than in the cases of SPS scenarios.  
This enhancement shows up for large $|A_b|$ and also near threshold 
regions.
In this scenario, the SUSY-QCD corrections are about 2\%,
the electroweak corrections are about 5\% while the QED corrections
are very small. The dominant contribution comes from the Yukawa
corrections and is enhanced by large $\tan\beta=60$ and large 
$|A_b|$.\\
In the left plot of Fig.(\ref{mssm}), the region $|A_b|=|A_t|<300$ GeV 
has no
data. This is due to the fact that splitting between $\wt{t}_2$ and
$\wt{t}_1$ ($\wt{t}_2$ and $\wt{b}_1$) is not large enough to allow 
the decays  $\wt{t}_2 \to \wt{t}_1 Z$
and $\wt{t}_2 \to \wt{b}_1 W$.\\
In the right plot of Fig.(\ref{mssm}), when 
$|A_b|=|A_t|\approx 0$ GeV, the splitting between $\wt{b}_2$ and
$\wt{t}_1$ is close to $m_W$ mass and so the tree level width for
$\wt{b}_2\to \wt{t}_1W^+$ 
almost vanish, consequently the correction is getting bigger.
This behavior has been also observed in previous plots for 
$\wt{b}_2\to \wt{t}_1W^+$.

\begin{figure}[t!]
\smallskip\smallskip 
\vskip-.5cm
\centerline{
{\epsfxsize3.2 in\epsffile{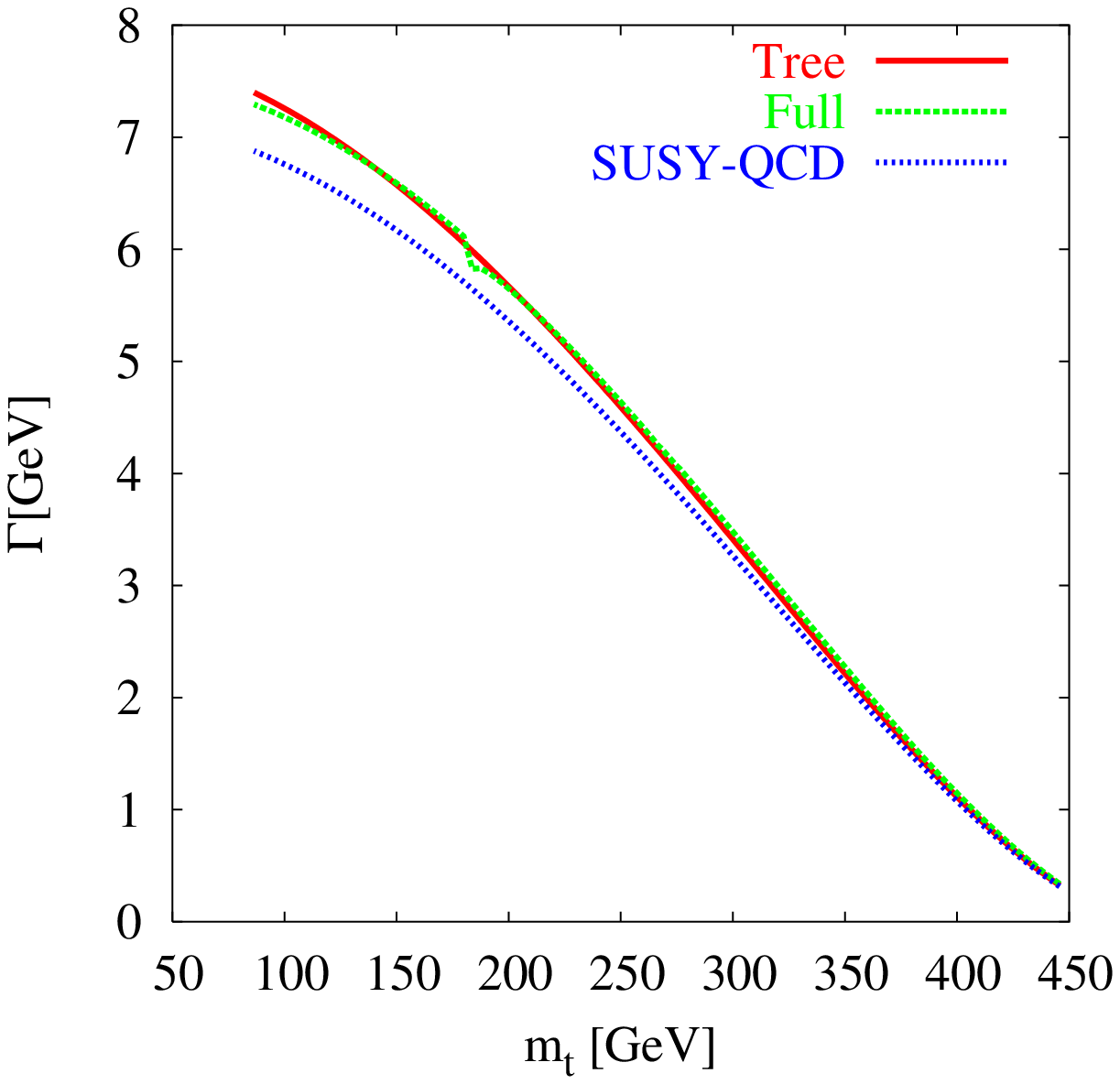}}  
\hskip-1.99cm
{\epsfxsize3.2 in\epsffile{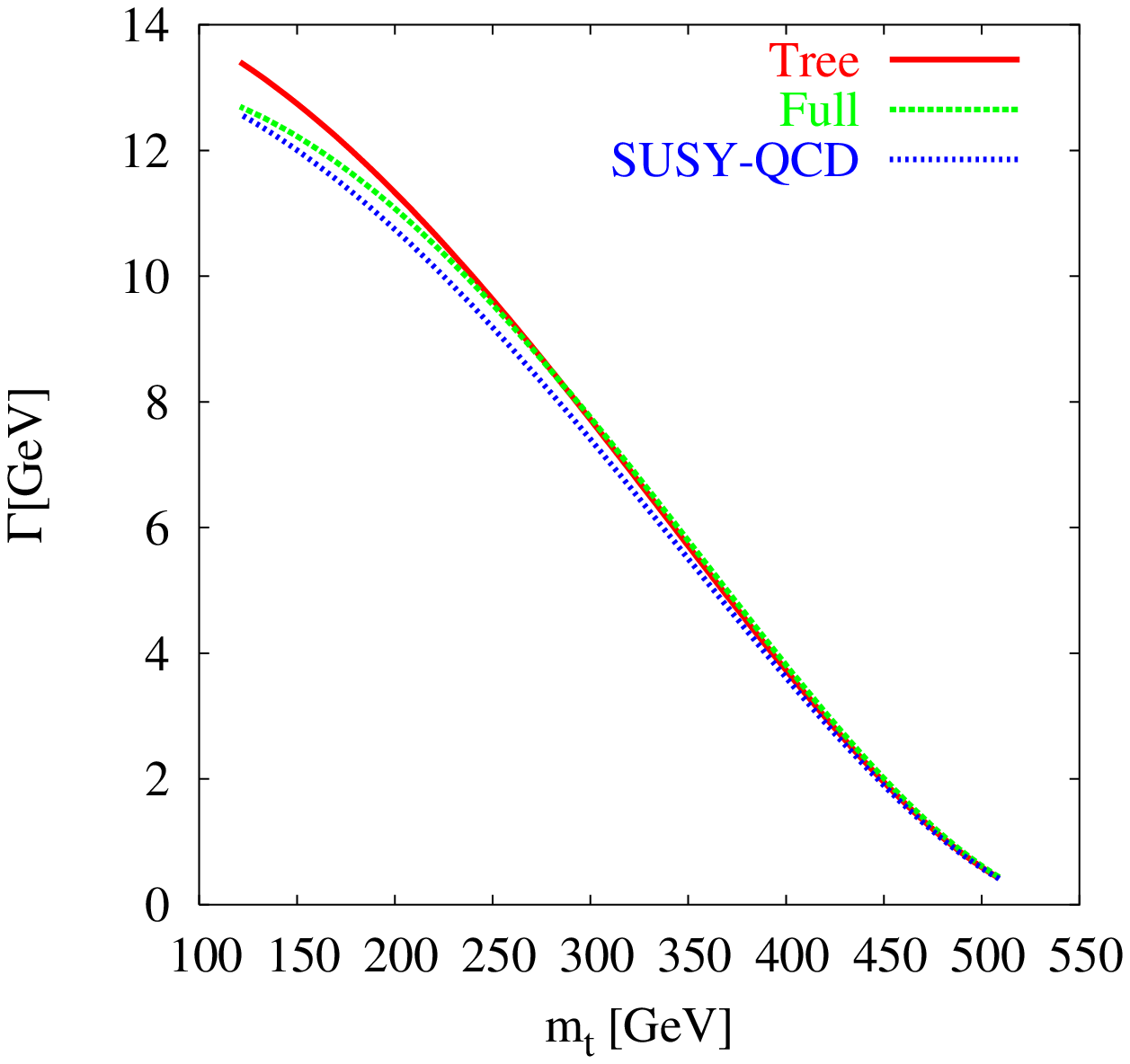}}}
\smallskip\smallskip
\caption{Tree and one loop decay width of $\wt{t}_2 \to \wt{t}_1 Z$ 
as function of $m_{\wt{t}_1}$}
\label{fig6}
\end{figure}

Finally, in Fig.~(\ref{fig6}) we illustrate the decay width of 
$\wt{t}_2 \to \wt{t}_1 Z$ as function of $m_{\wt{t}_1}$ 
in SPS1 (left) and  SPS5 (right). In SPS1
(resp SPS5) the decay width of $\wt{t}_2 \to \wt{t}_1 Z$ 
is about 8 GeV (resp 13 GeV)
for light stop mass of the order 100 GeV. 
Obviously, these decays width decrease 
as the light stop mass increase.

It is clear that the SUSY-QCD corrections reduces the width 
while the electroweak
corrections cancel part of those QCD corrections.
Both in SPS1 and SPS5, the full one loop width of 
$\wt{t}_2 \to \wt{t}_1 Z$  is in some case slightly bigger 
than the tree level width.\\

\noindent
{\bf 4.} To conclude, a full one-loop calculations
of third-generation scalar-fermion decays into gauge bosons W and Z
are presented in the on--shell scheme. 
We include both electroweak, QED and SUSY-QCD
contributions to the decay width. It is found that 
the electroweak and SUSY-QCD corrections 
interfere destructively.\\
The size of the one-loop effects are 
typically of the order $-5 \%\to 10$ \% in SPS scenarios which are 
based on
SUGRA assumptions. While in the general MSSM, 
the size of the corrections are bigger and can reach 
about 20\% for large $\tan\beta$ and large soft SUSY breaking $A_b$.
Their inclusion in phenomenological studies and analyses are
then well motivated.

\section*{{{Acknowledgment:}}} 
We are very grateful to Prof. Mohamed Chabab and the Organizing
committee for the invitation to ICHEMP05 and for the kind
hospitality at Cadi Ayyad Uninversity.
A.A is supported by the Physics Division of
National Center for Theoretical Sciences under a grant from the
National Science Council of Taiwan.
This work is supported by PROTARS-III D16/04.


\begin{thebibliography}{99}
\bibitem{CDF}
T.~Affolder {\it et al,}
Phys.\ Rev.\ D {\bf 63}, 091101 (2001);
G.~Abbiendi {\it et al,}
Phys.\ Lett.\ B {\bf 545}, 272 (2002)
[Erratum-ibid.\ B {\bf 548}, 258 (2002)];


\bibitem{LC}
J.~A.~Aguilar-Saavedra {\it et al},
arXiv:hep-ph/0106315;
K.~Abe {\it et al},
arXiv:hep-ph/0109166;
T.~Abe {\it et al},
arXiv:hep-ex/0106056.


\bibitem{gh1}
A.~Djouadi, W.~Hollik and C.~Junger,
Phys.\ Rev.\ D {\bf 55}, 6975 (1997);
S.~Kraml, H.~Eberl, A.~Bartl, W.~Majerotto and W.~Porod,
Phys.\ Lett.\ B {\bf 386}, 175 (1996).

\bibitem{gh2} J.~Guasch, W.~Hollik and J.~Sola,
order,''
JHEP {\bf 0210}, 040 (2002);
J.~Guasch, J.~Sola and W.~Hollik,
Phys.\ Lett.\ B {\bf 437}, 88 (1998);


\bibitem{DHAJ}
A.~Arhrib, A.~Djouadi, W.~Hollik and C.~Junger,
Phys.\ Rev.\ D {\bf 57}, 5860 (1998).


\bibitem{9701336}
A.~Bartl et al,
Z.\ Phys.\ C {\bf 76}, 549 (1997).

\bibitem{ah}
A.~Arhrib and W.~Hollik,
  JHEP {\bf 0404}, 073 (2004).

\bibitem{SPS}
B.~C.~Allanach {\it et al.},
Eur.\ Phys.\ J.\ C {\bf 25}, 113 (2002);
N.~Ghodbane and H.~U.~Martyn, arXiv:hep-ph/0201233.

\bibitem{ab}
A.~Arhrib and R.~Benbrik,
  Phys.\ Rev.\ D {\bf 71}, 095001 (2005).

\bibitem{qcdv} 
A.~Bartl et al
bosons,''
Phys.\ Lett.\ B {\bf 419}, 243 (1998).

\bibitem{sdecay1}
A.~Djouadi, J.~L.~Kneur and G.~Moultaka,
arXiv:hep-ph/0211331.
\bibitem{sdecay2}
M.~Muhlleitner, A.~Djouadi and Y.~Mambrini,
particles in the
arXiv:hep-ph/0311167.


\bibitem{FA} J.~Kublbeck, M.~Bohm, A.~Denner,
Comput.\ Phys.\ Commun.\  {\bf 60}, 165 (1990);
T.~Hahn, Comput.\ Phys.\ Commun.\  {\bf 140}, 418 (2001);
T.~Hahn, C.~Schappacher,
Comput.\ Phys.\ Commun.\  {\bf 143}, 54 (2002);
T.~Hahn et al,
Comput.\ Phys.\ Commun.\  {\bf 118}, 153 (1999);

\bibitem{FF} G.~J.~van Oldenborgh,
Comput.\ Phys.\ Commun.\  {\bf 66}, 1 (1991);
T.~Hahn, Acta Phys.\ Polon.\ B {\bf 30}, 3469 (1999)


\bibitem{Denner}
A.~Denner,
Lep-200,''
Fortsch.\ Phys.\  {\bf 41}, 307 (1993).


\bibitem{ren}
W.~Hollik et al,
Nucl.\ Phys.\ B {\bf 639}, 3 (2002);
W.~Majerotto,
arXiv:hep-ph/0209137;
T.~Fritzsche and W. Hollik,
Eur.\ Phys.\ J.\ C {\bf 24}, 619 (2002);

\bibitem{Eidelman}
S.~Eidelman and F.~Jegerlehner, {\em Z. Phys.} {\bf C67} (1995)
585--602.
\bibitem{susyqcd} 
A.~Arhrib, M.~Capdequi-Peyranere and A.~Djouadi,
  Phys.\ Rev.\ D {\bf 52}, 1404 (1995).
H.~Eberl, A.~Bartl and W.~Majerotto,
  Nucl.\ Phys.\ B {\bf 472}, 481 (1996).
\end{thebibliography}
\end{document}